\documentclass[prx,aps,amssymb,amsmath,amsmath,superscriptaddress,reprint,twocolumn,footinbib]{revtex4-2}
\usepackage{graphicx}
\usepackage{epstopdf}
\usepackage{natbib}
\usepackage{hyperref}
\usepackage{dcolumn}
\usepackage{bm}
\usepackage{dcolumn}
\usepackage{multirow}

\usepackage{graphicx}
\usepackage{amssymb}
\usepackage{tabularx}
\usepackage{amsmath}

\usepackage{xcolor}

\def\cO{{\mathcal O}}

\def\11{{\mathbb 1}}

\def\GeV{\,\rm GeV}

\def\cd{\!\cdot\!}

\def\beq{\begin{equation}}
\def\eeq{\end{equation}}
\def\bea{\begin{eqnarray}}
\def\eea{\end{eqnarray}}
\def\nn{\nonumber}

\newcolumntype{d}[1]{D{.}{.}{#1}}

\newcommand{\Fkt}[1]{\,\mathsf {#1}}
\def\openone{\leavevmode\hbox{\small1\kern-3.3pt\normalsize1}}

\ifx\Tr\renewcommand{\Tr}{\Fkt{Tr}} 
\else\newcommand{\Tr}{\Fkt{Tr}}
\fi

\usepackage{booktabs}
\usepackage{graphicx}
\usepackage{amsmath}
\usepackage{indentfirst}
\begin{document}

\title{Signatures of supermassive charged gravitinos\\[1
mm]  in liquid scintillator detectors}

\author{\sc Adrianna Kruk}
\affiliation{\sl Faculty of Chemistry, University of Warsaw, Pasteura 1, 02-093 Warsaw, Poland}
\author{\sc Micha\l\ Lesiuk}
\affiliation{\sl Faculty of Chemistry, University of Warsaw, Pasteura 1, 02-093 Warsaw, Poland}
\author{\sc Krzysztof A. Meissner}
\affiliation{\sl Faculty of Physics, University of Warsaw, Pasteura 5, 02-093 Warsaw, Poland}
\author{\sc Hermann Nicolai}
\affiliation{\sl Max-Planck-Institut f\"ur Gravitationsphysik (Albert-Einstein-Institut) \\
M\"uhlenberg 1, D-14476 Potsdam, Germany}
\vspace{1.0cm}

\date{\today}

\begin{abstract}
In a previous work [K.A. Meissner and H. Nicolai, Eur. Phys. J. C {\bf 84}, 269 (2024)], two of the present authors  have suggested possible experimental ways to search for stable supermassive particles with electric charges of $\cO(1)$ in upcoming underground experiments, in particular the new Jiangmen Underground Neutrino Observatory (JUNO) experiment. In the current paper, we present a detailed analysis of the specific signature of such gravitino-induced events for the JUNO detector and for upcoming liquid argon detectors like DUNE (Deep Underground Neutrino Experiment). The proposed method of detection relies on the ``glow'' produced by photons during the passage of such particles through the detector liquid, which would last for about a few to a few hundred microseconds depending on its velocity and the track. The cross sections for electronic excitation of the main component of the scintillator liquid, namely linear alkylbenzene (LAB), by the passing gravitino are evaluated using quantum-chemical methods. The results show that, if such particles exist, the resulting signals would lead to a unique and unmistakable signature, for which we present event simulations as they would be seen by the JUNO or DUNE photomultipliers. Our analysis brings together two very different research areas, namely fundamental particles physics and the search for a fundamental theory on the one hand, and methods of advanced quantum chemistry on the other.
\end{abstract}

\maketitle

\section{Introduction}
\label{sec:intro}

Following the suggestion \cite{MeissnerNicolai2019} that Dark Matter 
(or DM, for short) could consist at least in part of an extremely dilute 
gas  of supermassive stable gravitinos with charge $\pm\frac23 e$, 
possible experimental signatures of the passage of such particles
through underground detectors were examined in \cite{MN} (see also 
\cite{MN0} where an outlier event reported by the MACRO collaboration
\cite{MACRO1} was investigated as possible evidence towards 
the existence of such a new particle).
The suggestion of superheavy gravitinos as DM is chiefly motivated
by the remarkable match between the fermion content of the Standard
Model (SM) of particle physics with three generations of 16 quarks and leptons 
on the one hand, and the spin-$\frac12$ states of the $N\!=\!8$ supermultiplet
associated to maximally extended supergravity on the other 
hand \cite{GM} (see also \cite{NW}). This agreement and the
link with $N\!=\!8$ supergravity lead to the prediction 
that the only extra fermions beyond the observed 48 spin-$\frac12$ fermions 
of the SM would be eight supermassive gravitinos, which 
carry fractional electric charges \cite{MN2,KM0}. 

In this paper we analyze options for a dedicated search 
for slowly moving supermassive charged gravitinos in much more detail
than has been done before. We shall argue that the JUNO experiment \cite{JUNO}
is predestined for such a search, due to its special design and the 
properties of its scintillator liquid. JUNO is the largest underground detector ever 
constructed, with the main purpose of exploring the physics of (anti-)neutrinos 
(see \cite{JUNO,JUNOtesting} for a detailed description of the experiment), 
and expected to start taking data in early 2025,  this close date providing an extra incentive
for the present work. From our perspective, JUNO stands out among other
upcoming and planned such experiments because it offers unique opportunities 
to also search for exotic events of the type proposed here.
Specifically, we will focus on a particular proposal for embedding the SM into 
a Planck scale unified theory, and investigate its possible experimental
consequences. We will show that if supermassive particles with the properties
predicted by this scheme do exist, the associated signatures would be very different from 
those induced by neutrino events, and could thus be easily discriminated 
with appropriate (and hopefully rather minor) adjustments of the 
JUNO electronics and triggering systems. It goes without saying that 
the actual discovery of these still hypothetical particles would constitute a
major advance also in the search for a unified theory. Although we concentrate on JUNO here, our considerations apply equally well to upcoming new experiments based on liquid argon, like DUNE~\cite{DUNE}, for which we also present analogous theoretical estimates.

Our paper thus tries to predict, as accurately as possible and taking into account all 
relevant aspects, ways and means to search for weakly ionizing charged 
supermassive gravitinos with the JUNO detector. 
Remarkably, our analysis requires an interdisciplinary endeavor 
on the interface of particle physics and quantum chemistry, two areas 
which are usually considered to be far apart. The particle physics part of 
the paper involves a novel approach to physics beyond the SM
which we review in section II, and relies on a still to be developed theoretical
framework beyond quantum field theory, as well as mostly 
unexplored mathematics (hyperbolic Kac-Moody Lie algebras). 
The quantum chemistry part makes use of advanced methods and large
computational resources to calculate the lowest excitation levels of the 
scintillator oil. The latter are needed for the precise determination
of the emission rates of photons along the track of the gravitino in the detector, 
which in turn are crucial for the observability (or not) of such events.
As it turns out, the specifics of the JUNO scintillator liquid play a key role
in this computation. As a first step, the present analysis considers only the photons emitted 
from the lowest excitation levels of the scintillator oil. It thus gives only a 
{\em lower bound} on what could actually be seen, but this seems already 
sufficient to detect the gravitinos when they pass through the JUNO detector, provided the 
triggering and software are appropriately designed. In future work we will discuss 
further effects that may enhance the predicted signal (higher level excitations, 
ionization, {\em etc.}). We also explain the possible reasons why existing
WIMP searches using solid state or liquid Argon detectors were unlikely to detect any 
such signals -- not to mention the fact that the signals these experiments have been  
concentrating on are of an entirely different character! In the section on simulations we will also
list possible caveats and difficulties that may still stand in the way of discovery, 
such as the dark count rates of the JUNO photomultiplier tubes (PMTs), or the $^{14}$C 
contamination of the detector liquid  (whose concentration in the detector is bounded 
above by $2\cdot 10^{-17}$  according to most recent measurements \cite{Oberauer}).

Simulating the interactions between the scintillating medium and a passing supermassive
gravitino is a complicated task, in part due to the existence of several competing processes. 
Apart from direct interactions with nuclei (nuclear recoil), the three main 
mechanisms by which the energy can be transferred from a charged particle 
(projectile) to a molecule in a liquid are: (i) excitation to a bound excited state of the molecule, 
(ii) capture of an electron by the projectile and its subsequent re-absorption 
by a molecular ion, and (iii) ionization, {\em i.e.} ejection of electron to the 
continuum \cite{MN}.  In this work we 
will only discuss the first process. While the remaining two processes certainly contribute, we show 
that the first one alone may already lead to a signal which is strong enough to be detectable by
JUNO. Therefore the results of our work  should be considered as a very conservative {\em lower bound} 
on the signal strength, not a definitive quantitative prediction. In subsequent papers we shall 
consider the other mechanisms outlined above which may enhance the signal.

The organization of this paper is as follows. In Section~II we list the main reasons for the gravitino DM hypothesis, and its main differences with more established ideas about DM, namely the mass of the order of $10^{18}$ GeV and the $\cO(1)$ electric charge of the proposed DM candidate. In the following sections we first briefly review the JUNO detector design and then study the interaction of the gravitino with the liquid scintillator medium. This enables us to explain the mechanism by which a passing gravitino leads to emission of light which may be subsequently detected. Next, by taking into account other factors that play a role in the detection, we theoretically predict the signature of the signal that the gravitino would produce. We discuss how the signal from the gravitino would differ from other events that may occur within the detector, 
in particular the inverse beta decays which JUNO was designed to record (Section~IV). We argue that the signal produced by a gravitino passing through the detector is fairly unique and can be distinguished both from the background noise and, being much longer, from the inverse beta decay events or antineutrino interactions with the nuclei. Next, we use quantum-chemical methods to compute the relevant cross sections for electronic excitation caused by the gravitino, for both liquid argon detectors and JUNO. The case of argon is computationally much simpler and also potentially useful for possible future searches with such detectors. The details of the computational model and the quantum-chemical calculations are explained in Section~V. Finally, in Section~VI we present simulations describing how the JUNO PMTs could record a gravitino event. Here we provide results for the case of a gravitino with two assumed velocities corresponding the Earth velocity and the galactic velocity, respectively, to argue that even in this worse case of the Earth velocity the track produced by the gravitino can be unambiguously identified and clearly discriminated against background or neutrino induced events. Our simulations take into account background event rates due to the $^{14}$C contamination of the scintillator fluid as well as dark count rates of photomultipliers, both of which turn out to be much smaller (in the Earth velocity case) or negligible (in the galactic velocity case) than the signal. The final remarks and directions for future work are given in Section~VII.

\section{Why supermassive charged gravitinos?}

Because the proposed scenario differs substantially from more established ideas about DM, and for the reader's convenience, we first review  the main arguments for supermassive charged gravitinos and the gravitino 
DM hypothesis.  This proposal has its origins in maximal $N\!=\!8$ supergravity 
\cite{CJ,dWN} and relies in an essential way on the  
associated fermion multiplet comprising 56 helicity-$\frac12$ 
and eight helicity-$\frac32$ states (this supermultiplet is also distinguished 
by the cancellation of conformal anomalies \cite{ConfAnom,T}).
In comparison with currently popular 
proposals for extending the SM, which usually come with numerous
new particles not too far above the electroweak scale (especially in 
the context of superstring unification and low energy supersymmetry), 
this scheme is very minimalistic in that the only extra fermions appearing 
beyond the 48 spin-$\frac12$ fermions of the SM are eight supermassive 
gravitinos (which absorb eight spin-$\frac12$ Goldstinos).
Such a novel approach to unification and the DM problem appears 
amply justified in view of decades of  failed attempts to discover 
any signs of new physics with more standard approaches
to physics beyond the SM, most significantly the non-observation 
of new fundamental spin-$\frac12$ degrees of freedom 
at the Large Hadron Collider (LHC). A further hint in support of a minimalistic scheme
is the fact that the renormalization group (RG) evolution of the measured SM couplings up 
to the Planck scale exhibits no Landau poles (an apparent instability 
in the effective scalar potential generated by the RG evolution of the Higgs 
boson self-coupling can be cured by slightly modifying the scalar sector of the SM).
This strongly indicates that, as far as its spin-$\frac12$ and spin-1 sectors are concerned,
the SM could simply survive more or less {\em as is} all the way up to the Planck scale --
an option that is outside the scope of conventional Grand Unified Theory (GUT) or superstring phenomenology.

In essence, the proposal 
goes back to Gell-Mann's old observation that the fermion content of the SM
with three, and only three, generations of quarks and leptons (including
right-chiral neutrinos) can be matched with the spin-$\frac12$ content of the
maximal $N\!=\!8$ supermultiplet after removal of eight Goldstinos \cite{GM,NW}.
This surprising agreement constitutes the main theoretical motivation 
for the present proposal. Yet, it has not received much attention because
of an overwhelming expectation  that LHC and related experiments, as well as
various DM searches, would discover, or at least produce indirect evidence for
new fundamental degrees of freedom at the TeV scale, especially in connection with low energy 
($N\!=\!1$) supersymmetry. The fact that all these searches have come up empty-handed 
despite enormous efforts strengthens the case for unification attempts such as 
the present one that try to make do with just the 48 SM spin-$\frac12$ fermions.

Crucially, in Gell-Mann's scheme the matching of U(1)$_{em}$ charges requires 
a `spurion shift' of $\delta q = \pm \frac16$ \cite{GM,NW} that is not part of $N\!=\!8$ 
supergravity and that, in terms of the original spin-$\frac12$  tri-spinor $\chi^{ijk}$ 
of  $N\!=\!8$ supergravity, requires a very peculiar deformation of the supergravity 
U(1) with a generator that does not belong to the SU(8) $R$-symmetry of the 
theory \cite{MN2,KM0}.  Furthermore, it invokes a fictitious family 
symmetry SU(3)$_f$ such that the supergravity fermions can only 
`see'  the diagonal subgroup of SU(3)$_c$ and SU(3)$_f$. The main
new step taken in \cite{KM0,MeissnerNicolai2019} consisted in extending 
Gell-Mann's considerations to the eight massive gravitinos which would be 
the only extra fermions  beyond the known three generations of quarks
and leptons. Under SU(3)$\,\times\,$U(1)$_{em}$ the gravitinos transform as 
\beq\label{GravCharges}
         \left({\bf 3}\,,\,\frac13\right) \oplus \left(\bar{\bf 3}\,,\,-\frac13\right)
         \oplus \left({\bf 1}\,,\,\frac23\right) \oplus \left({\bf 1}\,,\, -\frac23\right)
\eeq
Importantly, the U(1)$_{em}$ gravitino charge assignments in (\ref{GravCharges})
include the spurion shift needed for matching the spin-$\frac12$ 
sectors \cite{MeissnerNicolai2019}; see  \cite{MN2,KM0} for a derivation
of these quantum numbers and an explanation of how to extend the 
spurion shift  to the gravitinos. All gravitinos would thus carry 
fractional electric charges. If one identifies the SU(3) in (\ref{GravCharges}) 
with SU(3)$_c$ as in \cite{MeissnerNicolai2019}, a complex triplet of gravitinos would 
be subject to strong interactions.  The color triplet gravitinos (whose possible
physical manifestations have been discussed in \cite{MN3}) will not play 
any role in our subsequent considerations, so we will be concerned 
here only with the color singlet gravitinos of charge $\pm \frac23 e$. 

Although the combined spin-$\frac12$ and  spin-$\frac32$ content 
thus coincides with the fermionic part of the $N\!=\!8$ supergravity 
multiplet, it is important to stress that the underlying theory would 
need to be an as yet unknown extension beyond  $N\!=\!8$ 
supergravity, because the latter theory cannot accommodate the required
shift of U(1) charges, nor reproduce the SU(3)$_c \times$ SU(2)$_w\times$ 
U(1)$_Y$ symmetry of the SM. The point of view taken here is that the required 
extension necessitates  a framework {\em beyond} space-time based 
quantum field theory, and that rectification of the mismatches should 
be viewed as providing guidance towards such a theory. What seems 
clear is that such a theory will require extending known symmetries 
to infinite-dimensional symmetries, which may not admit a space-time realization. 
In particular, it is expected that the maximal rank hyperbolic Kac-Moody 
algebra E$_{10}$ and its `maximal compact'  sub-algebra K(E$_{10}$) 
will play a key role~\footnote{On the purely mathematical side not much is known
 about the Kac-Moody Lie algebra E$_{10}$, apart from its mere existence. This is
 due to the fact that this algebra, like all algebras of this type, exhibits 
 an exponential growth such that it is not even possible to give
 a complete list of its generators; {\em idem} for its involutory subalgebra 
K(E$_{10}$).}, and possibly supersede supersymmetry as a guiding 
principle towards unification. One strong hint in this direction is the fact that 
the requisite shift  of U(1) charges necessarily requires enlarging the $R$-symmetry 
SU(8)$\,\equiv\,$ K(E$_7$) of $N\!=\!8$ supergravity all the way to 
K(E$_{10}$) \cite{KN}, which is hugely infinite-dimensional.
At this time it is not clear how to go about constructing such a theory, but 
see \cite{DHN} for a closely related attempt in the context of 
maximal supergravity. From that construction it follows that conventional concepts 
of quantum field theory, as well as the SM symmetries and space-time 
itself,  would then have to {\em emerge} from some pre-geometric 
`space-time-less' scheme~\footnote{It is a very peculiar fact that K(E$_{10}$), despite being
infinite-dimensional, admits {\em finite-dimensional} (unfaithful) representations,
one of which corresponds to the fermionic part of the $N\!=\!8$ supermultiplet, 
hence also admitting a realization of the U(1) shift on these fermions (see \cite{KN}
and references therein). By contrast, this symmetry, and thus the U(1) deformation, 
require an infinite-dimensional representation for the bosons, thus involving 
new degrees of freedom for which no physical interpretation is currently known.
This is one reason for the distinguished role of fermions in the present context.}

\section{Quantitative estimates}

In view of the unresolved theoretical challenges we here adopt
a more pragmatic approach by focusing on possible 
observable signatures induced by the spin-$\frac32$ 
gravitinos which could support the above scenario. Accordingly, in this section we present some preliminary
estimates for various parameters to guide this search.
 Let us emphasize, however,  that at this point it is difficult to make  
quantitatively more precise predictions beyond giving plausible
ranges of various parameters, because several parameters, 
such as (\ref{gravmass}) and (\ref{flux}) below, remain unknown. Consequently,
our primary concern is to ascertain whether or not this completely new
kind of particle exists. In this regard the situation is analogous to the search for 
low energy ($N\!=\! 1$) supersymmetry, where -- even assuming that it is realized
in Nature at all -- there is still no firm prediction  as to when and where 
it should show up, despite an enormous  body of theoretical work, and
taking into account that large ranges of parameter values have
already been excluded by experiment.

An  important  consequence  of (\ref{GravCharges}) and the U(1) charge shift is 
that, due to their fractional charges the gravitinos cannot decay into SM fermions, and 
are therefore stable independently of their mass. Their stability against 
decays makes them natural candidates for DM \cite{MeissnerNicolai2019}.
While most currently popular DM scenarios are based on the assumption 
of ultralight constituents (such as axions) or TeV scale WIMPs, the above arguments
would suggest that DM could thus consist at least in part of an extremely dilute 
gas  of supermassive stable gravitinos with charge $q=\pm\frac23$ in units of the
elementary charge $e$. In this case they  must have
very large mass, as follows directly from the bounds on the electric 
charge of  any putative DM particle of large mass derived in \cite{milli1,milli2,NNP}.
Although these bounds were derived with a different perspective, namely 
bounding the charge of `low mass' (in comparison 
with the Planck scale) DM candidates, by considering various astrophysical
constraints, we are of course free to take the opposite view of this analysis as 
providing a {\em lower} bound on the mass of supermassive particles with 
given $\cO(1)$ charges. This evidently constitutes an extrapolation
well beyond the mass range considered in \cite{milli1,milli2,NNP},
but the inequalities derived there nevertheless indicate
that gravitinos with charges (\ref{GravCharges}) can be within the 
allowed range to be viable DM candidates. Because mutual
annihilation of gravitinos is not an issue due to their low abundance 
and extremely small annihilation cross section, and because elastic
scattering of Cosmic Microwave Background (CMB) photons off gravitinos has no noticeable effect 
on the CMB, the constraints are actually much less 
stringent than the ones obtained in those papers.
From the more recent  work~\cite{medvedev24} we can extract a lower bound
to arrive at the estimated range of mass values
\beq\label{gravmass}
10^{16} \, {\rm GeV} \,\lesssim \, m_{gravitino} \, \lesssim  10^{18}\, {\rm GeV}
\eeq
The lower bound is also in accord with the bound that would be obtained by
demanding that the gravitational attraction between two like-charged
gravitinos can overcome the electrostatic repulsion
between them, so as to enable gravitational collapse of gravitino lumps 
in the early radiation period for the formation of primordial black holes~\cite{meissner20}.
By contrast, the upper bound is motivated purely theoretically by  
demanding that the gravitino Schwarzschild radius  should not exceed 
its Compton wave length, and by the fact that `particle' and `mass'
may become meaningless notions beyond the Planck scale, in a 
regime where classical  space and time de-emerge.

As was argued in \cite{MeissnerNicolai2019} the abundance of the 
color singlet gravitinos cannot be estimated since they were never in thermal 
equilibrium, but if they are the dominant contribution to the DM content of the
universe one can plausibly assume their abundance in first approximation
to be given by the average DM density inside galaxies \cite{WdB}, 
which gives an upper bound
\beq\label{DM}
\rho_{DM} \,\lesssim \,0.3\cdot 10^6 \GeV\cd m^{-3}  
\eeq 
(the average in the Universe is a million times smaller). 
If DM were entirely made out of nearly Planck mass particles with an assumed 
mass of $10^{18}$ GeV, this would amount to  $\sim 3\cdot 10^{-13}$ 
particles per cubic meter within galaxies. Furthermore assuming an 
average velocity of 30 km$\,\cdot\,$s$^{-1}$ this yields the flux estimate
\beq\label{flux}
\Phi \;\sim \; 0.03 \, {\rm m}^{-2} {\rm yr}^{-1} {\rm sr}^{-1}
\eeq
based on the upper bound in (\ref{gravmass}); if the
gravitino mass is smaller this value could go up.
We emphasize that, in addition to the unknown mass of the particle, 
there remain important uncertainties which could shift this estimate  in
either direction. One concerns possible inhomogeneities in the DM distribution 
within galaxies or stellar systems, which could lead to either a local 
depletion or to a local enhancement of the DM density (\ref{DM}) 
in the vicinity of the earth (see {\em e.g.} \cite{DK}
for an early discussion of this point). The other is the average velocity $v$ 
of  superheavy DM particles w.r.t. the earth which we assume
to lie in the range 
\beq\label{v}
v_E \,\lesssim \, v \, \lesssim \,  v_S\, , 
\eeq
where $v_E$ is the virial velocity for a particle bound to the Sun and $v_S$ if it is bound to our Galaxy.
In the paper we will assume $v_E \sim 30$ km/s and $v_S \sim 230$ km/s. It is also clear 
that a charged particle with mass in the range (\ref{gravmass}) can easily pass through the earth 
without deflection, independently of the effects it has on ordinary matter. In particular, it can enter a 
detector from any direction. Because the velocities (\ref{v}) are
{\em non-relativistic} and the energy transfer from the particle to the molecule 
is on average below its ionization energy, ionization is a relatively rare event,
which is our main reason for concentrating on electronic excitations to bound states 
within the molecule. The low energy transfer is another  reason
why the lower bound in (\ref{gravmass}) less stringent than one would
infer from \cite{milli1,milli2,NNP}.

The distinctive feature of the present proposal is therefore that the DM gravitinos {\em do} 
participate in SM interactions with couplings of order $\cO(1)$. In this sense,
these DM candidates are not dark at all, but simply too rare and too faint to 
be observed directly with standard devices because of their extremely low 
abundance evident from (\ref{DM}).
This is in contrast to other scenarios involving DM particles which 
are assumed  to have only weak and gravitational interactions with SM matter,
but might still be seen with standard detectors due to their
more frequent occurrence. By contrast the gravitinos in (\ref{GravCharges})  
can in principle be directly detected thanks to their $\cO(1)$ SM interactions,
even taking into account the mentioned uncertainties regarding 
the estimates (\ref{gravmass}), (\ref{DM}) and (\ref{flux}).
A further obstacle is the fact that the rate of energy transfer from the gravitino to the scintillating medium is low, as explained above. It is here
that another special feature of JUNO that may facilitate
detection comes into play: the electronically-excited energy levels of the molecules in the scintillating liquid lie much lower in comparison with other media such as liquid argon. As a result, the excitation cross-section can be sufficiently large to produce a visible effect in the detector, despite the small overall energy transfer.

\section{Juno detector design}
\label{sec:juno}

For the benefit of the readers who are not familiar with the design of liquid scintillator detectors, in this section we provide a brief overview of salient features of the JUNO detector design and operation principles. First, JUNO is tuned to detect signals from the inverse beta decay events in which an antineutrino $\bar{\nu}_e$ hits a proton in the detector medium and produces a neutron and a positron
\begin{align}
    \bar{\nu}_e + p \longrightarrow n + e^+.
\end{align}
The positron annihilates almost immediately when it encounters an electron and produces two $\gamma$ photons 
-- the primary signal, and the  main event JUNO will be analyzing. Being an isolated
event, this signal is easy to distinguish from the gravitino signature.
A secondary signal is produced some time later by the liberated neutron. Let us
explain why this secondary process also does not interfere with the proposed gravitino signal, see
for instance Ref.~\cite{giammarchi14} and references therein
for the numbers quoted below.

For the liberated neutron, there are different options. It can come either from the proton from H or from $^{12}$C (and then we are left with a $^{11}$B nucleus which is stable). 
In either case the free neutron can then be captured by H to give a deuterium nucleus in an excited state which emits a 2.22 MeV photon 
producing a flash with an average delay of 260 ms \cite{giammarchi14}. This is an isolated event, always coming 
in connection with the two gammas from the earlier annihilation of $e^+$. 
Alternatively, the neutron can be captured by a $^{12}$C nucleus to give a $^{13}$C nucleus which is stable. The neutron has very low probability of not  being captured because of the large cross section of being either absorbed or scattered. If, on the other hand, the neutron sticks with the original carbon nucleus we get a $^{12}$B nucleus with a 
lifetime of 20 ms. Its decay produces an $e^-$ and an anti-neutrino with energies up to 14 MeV, where the range of electrons is at most 20 cm, which is again easily distinguishable from our gravitino. The crucial point here is that all these neutron induced processes are {\em isolated} events tied to the initial $2\gamma$ flash
which look nothing like the {\em quasi-continuous stream} of photons that we predict for the gravitino
(see Section VII below). The gravitino signal can thus be easily discriminated, 
independently of how much time the neutron spends in the scintillator liquid after its liberation. 
Consequently neutron capture cannot mask the signature we predict for a passing gravitino.

The JUNO is built deep underground to shield it from the cosmic radiation and consists of two major parts. The first is the central detector which is spherical in shape and has a diameter of $35.5\,$m. It is filled with the scintillating liquid comprising three ingredients: linear alkylbenzenes (LAB, solvent), 2,5-diphenyloxazole (PPO, $3\,$g/L) and 1,4-Bis(2-methylstyryl)benzol (Bis-MSB, 15$\,$mg/L). As the $\gamma$ photons produced in the events travel through the detector, they undergo Compton scattering and transfer some energy to the molecules of the scintillating liquid which thereby gets electronically excited. The LAB molecules, being the most abundant, absorb most of the energy of the passing photon. These molecules quickly transfer this energy to the main fluor (PPO) through direct collisions or non-radiative energy transfer mechanisms. The PPO molecule then rapidly emits a photon (wavelength $\approx 350\,$nm) upon de-excitation from the first excited to the ground electronic state. Finally, the Bis-MSB molecules serve as wavelength shifters. They absorb the photons emitted by PPO and, with a small delay, re-emit photons with a different wavelength. The purpose of this process is to shift the photon wavelength to values where the photon detectors are the most sensitive. The central detector is covered with photomultiplier tubes which capture the photons emitted inside. 

The second part of the JUNO detector is a cylindrical tank of diameter $\approx 38\,$m in which the central (spherical) detector is embedded. The tank is filled with ultrapure water and is used to detect the Cherenkov radiation emitted when high-energy particles, primarily muons, pass through. When the Cherenkov radiation is detected in the water tank, all photomutiplier tubes in the main detector which are on the path of the incoming particle are switched off. This procedure, called vetoing, is the main mechanism which enables to reduce the background noise resulting from, e.g. radioactivity of the surrounding rock and soil.

\section{Detection of the gravitino}
\label{sec:gdetection}

It is clear that the gravitino particle possesses different physical characteristics than the antineutrinos which the JUNO detector was designed to detect. As a result, the signal generated by the gravitino will be different from the inverse beta decay events in terms of its temporal characteristics. In this section we shall discuss the differences in timing of these two types of signals from the qualitative point of view, while in subsequent sections we present results of the calculations which enable to quantify the strength of the gravitino signal.

The first major difference in the gravitino detection is related to the vetoing mechanism. As discussed in Refs.~\cite{meissner20,MN} gravitino moves at much lower velocities than the background muons, much less than 1\% of the speed of light in vaccum. It is well known that the Cherenkov radiation is emitted only when a particle travels in a medium with a velocity larger than the speed of light in this medium. The speed of light in a medium is smaller by a factor of $1/n$ than in vacuum, where $n$ is the refractive index of the medium. For water (which fills the outer tank of the detector) the refractive index is roughly $n\approx 1.33$ which means that the speed of light in water is about 75\% of the speed in vacuum. This velocity is much larger than that of the gravitino which means that the passing gravitino will not emit the Cherenkov radiation. 

Nonetheless, the gravitino passing through the outer tank will excite the neighboring water molecules to higher electronic states. By subsequent radiative decay, water molecules may emit photons which are detected by the photomutiplier tubes. However, as the de-excitation of liquid water is dominated by non-radiative decay channels and the outer tank contains no fluor molecules, the efficiency of this process is extremely low. Moreover, the orientation of the photons potentially emitted by water molecules is completely random. This is different from the Cherenkov radiation which is emitted in a cone with the opening angle $\cos\theta = 1/n\beta$, where $\beta$ is the ratio between the speed of the particle and the speed of light in vacuum. For these two reasons, we argue that the passing gravitino will not trigger the vetoing mechanism in the outer tank.

Next, the passing gravitino enters the inner part of the detector filled with the scintillating mixture. It excites the surrounding LAB molecules which transfer the excitation to the PPO fluor through collisions. Light emitted by the radiative decay of the fluor is detected in the PMTs  (after absorption and re-emission by the wavelength shifter). On the surface, this mechanism is very similar to the inverse beta decay events where the LAB molecules are excited by Compton scattering, while the subsequent steps are identical. However, we argue that the inverse beta decay events can be distinguished from gravitino events based on their entirely different temporal characteristics.

The $\gamma$ photons generated by the inverse beta decay events travel trough the detector at roughly 66\% of speed of light in vacuum, because the refractive index of the scintillating liquid is about $n\approx 1.5$. As the inverse beta decay events can occur at an arbitrary point in the inner detector, the two $\gamma$ photons have to travel, on average, the distance equal to the radius of the detector, $R\approx 17.7\,$m. Therefore, it takes about $90\,$ns before the $\gamma$ photons leave the detector. All LAB molecules are excited within this time and it takes another couple of nanoseconds for the excitation transfer and decay of the PPO molecule. Therefore, the whole primary signal is over within roughly $100\,$ns. If the secondary signal resulting from the capture of the free neutron by atomic nucleus (and subsequent decay of the deuteron) is observed, another $\gamma$ photon is generated after about $250\,$ns from the primary event. The secondary signal is over after another $\approx 100\,$ns. In total, one would observe two strong signals lasting $\approx 100\,$ns separated by about $250\,$ns, and the whole event would last less than a microsecond.

The situation is quite different in the case of gravitinos. As mentioned above, they move at a much lower velocity, estimated at considerably less then 1\% of the speed of light. In fact, more rigorous velocity estimates were established in the previous section by assuming that DM particles are bound either to the
Solar System or to the Milky Way galaxy, leading to $v_E=29.8\,$km/s and $v_S=230\,$km/s, respectively. When expressed as a fraction of the speed of light, these velocities correspond to $\beta_E\approx10^{-4}$ and $\beta_S\approx7.7\cdot 10^{-4}$. This means that the gravitino needs either about $600\,\mu$s (for $v_E$) or about $77\,\mu$s (for $v_S$) to complete the run through the inner detector, assuming that the average distance it has to travel is equal to roughly the radius of the inner sphere. This corresponds to time scales larger by two or three orders of magnitude in comparison with the inverse beta decay events.

This difference in time scales can be used to distinguish the inverse beta decay events (and other events involving fast particles) from the gravitino events. Due to its enormous mass, the gravitino would move at nearly constant speed along a linear path. Every now and then, the LAB molecule is excited and this excitation is transferred to the PPO fluor which emits a photon. After, on average, roughly $100\,$ns from the LAB excitation, this photon reaches the photomultiplier tubes and can be detected. By this time the gravitino travels a certain distance and has a chance of exciting another LAB molecule, hence the process repeats itself. The signal will last for tens of microseconds up to a fraction of a milisecond, rather than less than a microsecond as in the inverse beta decay signals or the antineutrino interaction with nuclei. 

In the above analysis, we have assumed that the signal generated by the gravitino is strong enough to be observable by the JUNO detector. This depends entirely on the number of photons generated per unit distance as the gravitino travels through the scintillating fluid. To estimate this quantity, knowledge of the number of LAB molecules excited by the gravitino to a higher electronic state is required. In the next section, we develop a theoretical framework which enables to calculate this crucial parameter.

\section{Theory of molecular excitations}
\label{sec:theory}

\subsection{Preliminary considerations}
\label{subsec:initial}

As the interaction of the gravitino with the scintillating medium is a complicated process, approximations have to be introduced to make theoretical predictions feasible. In this section we introduce and justify the approximations introduced into the theoretical model. 

First, it is instructive to consider the time and energy scales involved in the interaction of the gravitino with a charge-neutral molecule. First, regarding the energy scale, in the preceding paper~\cite{MN} it has been discussed that in a simple classical picture, the maximum energy transferred from the gravitino to an electron in a molecule is of the order of several eV for $\beta\approx 10^{-3}$ or less. This corresponds to the energy range of 
electronic excitations to \emph{bound} excited states, whereas ionization events are expected to be relatively unimportant. In other words, the passing gravitino can be treated as a perturbation, similarly as with the interaction of weak external electric field. 

Of course, this perturbative picture is not valid in situations where the gravitino penetrates deep within the $1s$ core orbitals of carbon, oxygen, {\em etc.} atoms. In such cases, the excitation of the core electron leaves a vacancy filled by an electron from a higher energy level (Auger effect) and an ionization event is possible. However, such events are expected to be rare due to relatively compact extent of the core orbitals in comparison with the valence shells. For example, the volume of a single water molecule calculated as the volume of a union of spheres corresponding with the van der Waals radii of two hydrogen and an oxygen atoms equals to about $26$\AA$^3$~\cite{maglic22} (geometry of water in liquid taken from Ref.~\cite{moriarty97}). The effective radius of the water molecule is hence equal to about $1.8$\AA. By comparison, the $1s$ core orbital of the oxygen atom is characterized by the mean electron-nucleus distance of $\langle r\rangle = 0.105$\AA~\cite{saito09}. Due to the exponential decay of the orbital with the electron-nucleus distance, a sphere of radius $2\langle r\rangle = 0.210$\AA~centered at the atomic nucleus encloses nearly total core density ($\approx\,$94\%). The probability of the collision with the core relative to the collision with the whole molecule is equal to the square of the ratio of the corresponding effective radii and evaluates to only about $1$\%. For larger molecules, such as LAB, this ratio turns out to be even lower. Therefore, even if the core penetration events are not described accurately by our model, this should have a negligible effect on the overall results, justifying the perturbative approach. 

The second issue which warrants a discussion is the strength of the interaction between the molecules in a liquid in relation with the energy transfer from the gravitino. In the inner detector, interaction between LAB molecules is dominated by the $\pi$-$\sigma$ and $\pi$-$\pi$ dispersion interactions (London forces), because the permanent dipole moment of LAB is very small. The strength of this type of interactions is usually of the order of $10-15\,$kJ/mol, for example, $11.3\,$kJ/mol for the benzene dimer~\cite{pitonak08} and $14.5\,$kJ/mol for the toluene dimer~\cite{rogers06}. Therefore, the intermolecular interaction energies in the scintillating medium, on average, are smaller than $0.01\,$eV -- two to three orders of magnitude less than the expected energy transfer from the passing gravitino. This leads to the conclusion that it is reasonable to neglect the intermolecular interactions in modeling the excitations caused by the gravitino.

\begin{table}[t!]
    \caption{\label{tab:lab-composition} Components of LAB with general formula C$_6$H$_5$CHR$_1$R$_2$ and their molar concentrations.}    
    \begin{ruledtabular}
    \begin{tabular}{cccc}
    Overall formula & Concentration & R$_1$ & R$_2$ \\
    \hline
    \multirow{4}{*}{C$_6$H$_5$-C$_{10}$H$_{21}$} & 
    \multirow{4}{*}{10\%}
    & CH$_3$ & C$_8$H$_{17}$ \\
    & & C$_2$H$_5$ & C$_7$H$_{15}$ \\
    & & C$_3$H$_7$ & C$_6$H$_{13}$ \\
    & & C$_4$H$_9$ & C$_5$H$_{11}$ \\
    \hline 
    \multirow{5}{*}{C$_6$H$_5$-C$_{11}$H$_{23}$} & 
    \multirow{5}{*}{35\%}
    & CH$_3$ & C$_9$H$_{19}$ \\
    & & C$_2$H$_5$ & C$_8$H$_{17}$ \\
    & & C$_3$H$_7$ & C$_7$H$_{15}$ \\
    & & C$_4$H$_9$ & C$_6$H$_{13}$ \\
    & & C$_5$H$_{11}$ & C$_5$H$_{11}$ \\
    \hline 
    \multirow{5}{*}{C$_6$H$_5$-C$_{12}$H$_{25}$} & 
    \multirow{5}{*}{35\%}
    & CH$_3$ & C$_{10}$H$_{21}$ \\
    & & C$_2$H$_5$ & C$_9$H$_{19}$ \\
    & & C$_3$H$_7$ & C$_8$H$_{17}$ \\
    & & C$_4$H$_9$ & C$_7$H$_{15}$ \\
    & & C$_5$H$_{11}$ & C$_6$H$_{13}$ \\
    \hline
    \multirow{6}{*}{C$_6$H$_5$-C$_{13}$H$_{27}$} & 
    \multirow{6}{*}{20\%}
    & CH$_3$ & C$_{11}$H$_{23}$ \\
    & & C$_2$H$_5$ & C$_{10}$H$_{21}$ \\
    & & C$_3$H$_7$ & C$_9$H$_{19}$ \\
    & & C$_4$H$_9$ & C$_8$H$_{17}$ \\
    & & C$_5$H$_{11}$ & C$_7$H$_{15}$ \\
    & & C$_6$H$_{13}$ & C$_6$H$_{13}$ \\
    \end{tabular}
    \end{ruledtabular}
\end{table}

\begin{table}[t!]
    \caption{\label{tab:lab-excitation} Excitation energies and oscillator strengths for the $S_1$ state of the LAB isomers. The order of the isomers is the same as in Table~\ref{tab:lab-composition} (from top to bottom). The results for the proposed LAB model molecule (3-PP) are given in the last row, while averaged results for $20$ isomers are given in the second last row. All results were obtained at the ADC(2)/aug-cc-pVTZ level of theory using B3LYP-D3/cc-pVTZ optimized geometries.}    
    \begin{ruledtabular}
    \begin{tabular}{ccc}
    Isomer & Excitation energy (eV) & Osc. strength $\cdot\,10^{4}$ \\
    \hline
    1 & 5.195 & 6.49 \\
    2 & 5.184 & 7.33 \\
    3 & 5.183 & 7.25 \\
    4 & 5.182 & 8.20 \\
    \hline 
    5 & 5.194 & 6.35 \\
    6 & 5.185 & 7.25 \\
    7 & 5.183 & 7.35 \\
    8 & 5.182 & 7.55 \\
    9 & 5.183 & 8.12 \\
    \hline 
    10 & 5.194 & 6.31 \\
    11 & 5.185 & 7.37 \\
    12 & 5.183 & 7.11 \\
    13 & 5.182 & 7.45 \\
    14 & 5.182 & 7.58 \\
    \hline
    15 & 5.195 & 6.29 \\
    16 & 5.185 & 7.26 \\
    17 & 5.183 & 7.17 \\
    18 & 5.182 & 7.26 \\
    19 & 5.182 & 7.48 \\
    20 & 5.183 & 8.17 \\
    \hline
    average & 5.185 & 7.27 \\
    \hline
    3-PP & 5.186 & 7.68 \\
    \end{tabular}
    \end{ruledtabular}
\end{table}

Next, we consider the time scale related to the gravitino-molecule interaction. One can easily estimate that when the gravitino is, say, $10\,$\AA~from the molecular center, the strength of the resulting interaction is too small to cause any electronic excitation. Then assuming even the lowest velocity estimate of $\beta_E\approx10^{-4}$, it takes only a few tens of femtoseconds for the gravitino to travel from the initial distance of $10\,$\AA~from the molecule to the distance of $10\,$\AA~on the other side. Within this regime, energy transfer from the gravitino to the LAB molecule is possible, leading to electronic excitations. However, during such a short time the nuclei of a molecule can be considered as essentially static. This leads to the important conclusion that we can consider only adiabatic electronic excitations caused by the gravitino and treat atomic nuclei to be essentially motionless during the process. The subsequent nuclear relaxation and other transformations occur at much longer time scales and by that time, the gravitino is already far away from a molecule.

Finally, we have to introduce a model suitable for description of the gravitino. The proposed gravitino is a point-like particle of charge 
$\pm \frac23 e$ and enormous mass $Mc^2\approx 10^{18}\,$GeV. Therefore, the loss of the kinetic energy of the gravitino due to energy transfer to the scintillating medium (estimated at several eV per molecule, see above) is negligible. Therefore, one can assume that the gravitino moves along a predefined linear path with a constant velocity on the scale of the whole detector. Additionally, due to the huge mass, one can treat the gravitino simply as a source of a moving classical Coulomb potential, neglecting the quantum effects such as tunneling etc. In fact, a more rigorous treatment would be based on the Li\'enard–Wiechert potential which takes into account the retardation effects. However, the correction to the Coulomb potential resulting from the retardation is of the leading order $\beta^2$ which is negligible for the gravitino with the estimated $\beta\ll10^{-2}$.

\subsection{Model linear alkylbenzene molecule}
\label{subsec:labmodel}

The main component of the scintillating medium in the inner detector (LAB) is not a single molecule, but rather a mixture of various isomers with the general chemical formula C$_6$H$_5$-C$_n$H$_{2n+1}$ with $n=10-13$. In this work, we use that the composition of LAB given in Ref.~\cite{lombardi19} for the pilot distillation and stripping tests at Daya Bay
Neutrino Laboratory for their future use in the JUNO detector. In Table~\ref{tab:lab-composition} we provide a listing of the $20$ possible isomers of LAB identified by us based on Ref.~\cite{lombardi19} along with their stated overall concentrations.

Due to the large number of isomers and their considerable size, we seek a simpler model molecule that captures the essential characteristics of LAB. In particular, we focus on the first excited singlet state ($S_1$) of LAB. While the passing gravitino causes excitations also to higher excited electronic states, they are expected to be relatively unstable compared to $S_1$ and quickly decay non-radiatively due to non-adiabatic and collisional processes. Taking into consideration that the time required to transfer the energy from LAB to the PPO fluor is of the order of nanoseconds, we expect the $S_1$ state to be the primary carrier of the excess energy. Of course, the decay of higher excited electronic states of LAB undoubtedly leads to an increase population of the $S_1$ state, but neglect of such events can be seen as a conservative lower bound to the overall efficiency of the process. In a follow-up work, we shall consider excitations to higher electronic states, as well as other competing processes such as electron capture.

To study the variation of the properties of the $S_1$ state among the LAB isomers, we first optimized the ground-state geometry of each isomer using the B3LYP density functional~\cite{becke98,lyp88,vosko80} with the D3 dispersion correction~\cite{grimme10,grimme11,smith16} and the cc-pVTZ basis set~\cite{dunning89}.
These calculations were performed with the help of the {\sc QChem} program package~\cite{qchem}.  The optimized structures of each isomer are given in Supplemental Material~\cite{supp}. Next, the vertical excitation energy to the $S_1$ state, denoted $\omega_1$, and the corresponding oscillator strength of the dipole transition
\begin{align}
    f_{01} = \frac{4\omega_1}{3}|\pmb{\mu}_{01}|^2,
\end{align}
where $\pmb{\mu}_{01}$ is the dipole $S_0\rightarrow S_1$ transition moment, was calculated for each isomer using the algebraic-diagrammatic construction method~\cite{schirmer82,trofimov06,hattig02,dreuw15}, ADC(2), within the aug-cc-pVTZ basis set~\cite{kendall92}. In Table~\ref{tab:lab-excitation} we list the results obtained for every isomer. Clearly, the variation of the results among the isomers is tiny in the case of the excitation energies and amounts to no more than $0.02\,$eV. A larger spread is obtained in the case of the oscillator strengths, but even for this quantity the maximum deviation from the mean is of the order of $10\%$. From the chemical point of view, this similarity among LAB isomers is straightforward to understand. The excitation to the $S_1$ state is dominated by the $\pi\rightarrow\pi^*$ transition within the benzene ring. The alkyl substitution groups serve as weak electron donors, but do not alter the excitation character. We can exploit this property by restricting our consideration to a simpler model molecule that captures the essential properties of the $S_1$ state. For this purpose, we propose the 3-phenylpentane (3-PP) molecule with the chemical formula C$_6$H$_5$CH(C$_2$H$_5$)$_2$. In Table~\ref{tab:lab-excitation} we provide the excitation energy and the oscillator strength for the $S_0\rightarrow S_1$ transition in the 3-PP molecule calculated at the same level of theory as for the actual LAB constituents. As one can see, the properties of the 3-PP are very close to the average of the corresponding properties of LAB isomers.

\subsection{Excitation probability due to the passing gravitino}
\label{subsec:exgrav}

The main unknown quantity determining the possibility of detecting the proposed gravitino particle in the JUNO detector is the probability that the LAB molecules in the inner detector are electronically excited by the passing DM particle. To specify the model used to calculated this probability from first principles, we rely on the assumptions and estimates outlined in Sec.~\ref{subsec:initial}. First, we neglect the interactions between the molecules in the scintillating liquid and hence consider interaction of an isolated model LAB molecule (3-PP) with the gravitino. The gravitino is treated classically as a moving charge interacting with all electrons and nuclei in the molecule through the Coulomb potential
\begin{align}
\label{vdm}
    V_{\mathrm{DM}}(t) = -\sum_i \frac{q}{|\mathbf{r}_i - \mathbf{s}(t)|}
    + \sum_A \frac{q Z_A}{|\mathbf{R}_A - \mathbf{s}(t)|},
\end{align}
where the summation indices $i$ and $A$ run over all electrons and nuclei, respectively, in the molecule, $Z_A$ are the nuclear charges, and $\mathbf{s}(t)$ is the fixed path of the gravitino. An explicit parametrization of the latter quantity will be specified later, but for now it suffices to write $\mathbf{s}(t)=\mathbf{s}_0 + \mathbf{v}t$, where $\mathbf{v}$ is the velocity of the gravitino and $\mathbf{s}_0$ is the point at which the gravitino is the closest to the origin of the coordinate system. The origin can be chosen arbitrarily -- this choice has no influence on the final results after proper averaging over the path of the gravitino is performed. For simplicity, in the present work the nuclear position was chosen in the case of argon atom, while the nuclear center of charge was used for LAB model. In this parametrization, both in the $t\rightarrow\infty$ and $t\rightarrow-\infty$ limits the gravitino is infinitely far away from the atom/molecule. Conversely, for $t=0$ the grativino distance to the chosen origin $\mathbf{s}_0$ is minimal along the path.

Time evolution of the system is governed by the Schr\"odinger equation
\begin{align}
\label{tdse}
    i\partial_t \Psi(\mathbf{r},\mathbf{R},t) = \hat{H}\Psi(\mathbf{r},\mathbf{R},t),
\end{align}
where $\mathbf{r}$ and $\mathbf{R}$ is a shorthand notation for coordinates of all electrons and nuclei, respectively, in the system. The total Hamiltonian of the system is a sum of the electronic Hamiltonian for the isolated molecule $(H_0)$ and the interaction potential with the gravitino
\begin{align}
    \hat{H} = \hat{H}_0 + V_{\mathrm{DM}}(t).
\end{align}
As discussed in Sec.~\ref{subsec:initial} the effective time of the interaction between the gravitino and the molecule is small and nuclear motion can be neglected. Therefore, the total wavefunction is expanded into a linear combination of adiabatically excited electronic states
\begin{align}
    \Psi(\mathbf{r},\mathbf{R}_0,t) = \sum_{n=0}^\infty c_n(t)\,e^{-i E_n t}\,
    \psi_n(\mathbf{r},\mathbf{R}_0),
\end{align}
where $\mathbf{R}_0$ is the reference geometry of the molecule. The phase factor $e^{-i E_n t}$ in the above formula is arbitrary and was included for convenience. The states $\psi_n(\mathbf{r},\mathbf{R}_0)$ are eigenfunctions of the molecular Hamiltonian
\begin{align}
    \hat{H}_0\psi_n(\mathbf{r},\mathbf{R}_0) = E_n \psi_n(\mathbf{r},\mathbf{R}_0),
\end{align}
where $E_n\equiv E_n(\mathbf{R}_0)$ is the electronic energy of the state $n$. According to the arguments from Sec.~\ref{subsec:initial} we assume that the $V_{\mathrm{DM}}(t)$ can be treated as a perturbation and solve Eq.~(\ref{tdse}) using the time-dependent perturbation theory. We expand the coefficients $c_n(t)$ in the orders of the perturbation
\begin{align}
    c_n(t) = c_n^{(0)}(t) + c_n^{(1)}(t) + \ldots
\end{align}
and treat $\hat{H}_0$ and $V_{\mathrm{DM}}(t)$ as a zeroth-order and a first-order quantity, respectively. Additionally, the unperturbed system is assumed to reside in the ground electronic state, i.e. $c_n^{(0)}(t)=\delta_{n0}$. As in the first Born approximation, time derivative of the amplitudes are found from
\begin{align}
    \dot{c}_n^{(1)}(t) = e^{-i \omega_n t}\langle \psi_n | V_{\mathrm{DM}}(t) | \psi_0 \rangle.
\end{align}
We are interested in values of the coefficients $c_n(t)$ for $t\rightarrow\infty$, i.e. when the gravitino gets far away from the molecule. To this end, we integrate both sides of the above expression over time within the interval $t\in(-\infty,+\infty)$ and use the fact that $\lim_{t\rightarrow-\infty}c_n^{(1)}(t)=\delta_{0n}$. This leads to (for $n\geq1$)
\begin{align}
\label{cn1}
    c_n^{(1)} \equiv
    \lim_{t\rightarrow\infty} c_n^{(1)}(t) = 
    \int_{-\infty}^{\infty} dt\;
    e^{-i \omega_n t}\,\langle \psi_n| V_{\mathrm{DM}}(t) 
    |\psi_0 \rangle.
\end{align}
We are interested in the probability $p_{0\rightarrow n}$ that the molecule is found in the excited state $n$ when the gravitino gets far away, i.e. the perturbation is effectively switched off (adiabatically). It is given by the expression
\begin{align}
\label{p0n}
    p_{0\rightarrow n} = |c_n^{(1)}|^2.
\end{align}

The present method of calculation of the transition probabilities is based directly on Eq.~(\ref{cn1}). Evaluation of the key quantity, $\langle \psi_n| V_{\mathrm{DM}}(t) |\psi_0 \rangle$, requires two main ingredients. The first is the transition density matrix from the ground state to the $n$-th excited state. In this work, it was evaluated using the ADC(2) method employing a truncation scheme identical to Ref.~\cite{mester18}. The second ingredient are the matrix elements of the Coulomb potential $V_{\mathrm{DM}}(t)$ within the current atomic basis. As the Gaussian basis set are used exclusively in this work, the required integrals can be evaluated analytically~\cite{boys50,mcmurchie78} and {\sc LibInt2} library~\cite{Libint2} was employed for this purpose.

This leaves only the integration over time which is performed numerically on a uniform grid with spacing $\Delta t=0.1$. We verified that further decrease of $\Delta t$ leads to no appreciable changes in the overall results. The integral over time is approximated by truncating it according to $\int_{-\infty}^{\infty} dt \approx \int_{-T}^{T} dt$, and the value of $T$ is chosen adaptively (for each path separately) such that the calculated transition probabilities are stable to within threshold of $10^{-13}$. While this approach is straightforward is the case of argon and water, for LAB molecules a large $T$ was required to achieve convergence. In order to accelerate the calculations, whenever the gravitino is separated by more than $400\,$ bohr from the molecular center, the quantity $\langle \psi_n| V_{\mathrm{DM}}(t) |\psi_0 \rangle$ is evaluated using the multipole expansion of the potential $V_{\mathrm{DM}}(t)$ rather than by direct calculations. In the present work, the multipole expansion was truncated at quadrupole terms. This leads to an error vanishing as inverse fourth power of the gravitino-molecule separation which is negligible for the present purposes. Further technical details of the procedure of evaluating $p_{0\rightarrow n}$, along with analysis of additional approximations which can be applied to reduce the cost of the calculations, are beyond the scope of the present paper and will be reported separately.

\subsection{Gravitino path averaging}
\label{subsec:gpath}

\begin{figure}
    \centering
    \includegraphics[scale=1.0]{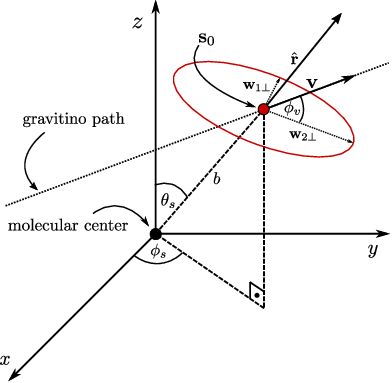}
    \caption{Parametrization of the path of the gravitino adopted in the present work.}
    \label{fig:path}
\end{figure}

To faithfully model the interaction of the gravitino with the scintillating medium, we have to take into account the fact that the gravitino encounters roughly $10^9$--$10^{12}$ molecules along its track which are randomly oriented with respect to the path. Therefore, the excitation probabilities $p_{0\rightarrow n}$ defined in the previous section have to be averaged over all possible path orientations.

As discussed in the previous section, the path of the gravitino is parametrized as $\mathbf{s}(t)=\mathbf{s}_0 + \mathbf{v}t$, where $\mathbf{s}_0$ is the point at which the gravitino is the closest to the adopted origin of the coordinate system. Therefore, there are five degrees of freedom in specifying the path -- one additional degree has already been eliminated by fixing the origin of the coordinate system (removing the translation of the system as a whole). Two variables that parameterize the path come up naturally in the present context. The first is the velocity of the gravitino, $v=|\mathbf{v}|$, and the second is the so-called impact parameter, $b=|\mathbf{s}_0|$, which is simply the minimal distance between the gravitino and the molecular center of charge along the path. Note that the definition of $\mathbf{s}_0$ implies that the dot product $\mathbf{v}\cdot\mathbf{s}_0$ is equal to zero.

This leaves three degrees of freedom to be specified. As the next two variables we choose angles $\theta_s\in[0,\pi]$ and $\phi_s\in[0,2\pi]$ which are the coordinates of the vector $\mathbf{s}_0$ in the spherical coordinate system centered at the origin, i.e. $\mathbf{s}_0=b\,[\sin\theta_s\cos\phi_s,\sin\theta_s\sin\phi_s,\cos\theta_s] = b \, \hat{\mathbf{s}}_0$, where $\hat{\mathbf{s}}_0$ is unit vector. For better clarity, we decided to express $\hat{\mathbf{s}}_0$ in Cartesian coordinates, $\hat{\mathbf{s}}_0 = [x_0, y_0, z_0]$, in the subsequent formulas.

\begin{figure*}
    \centering
    \begin{tabular}{cc}
    \includegraphics[scale=0.85]{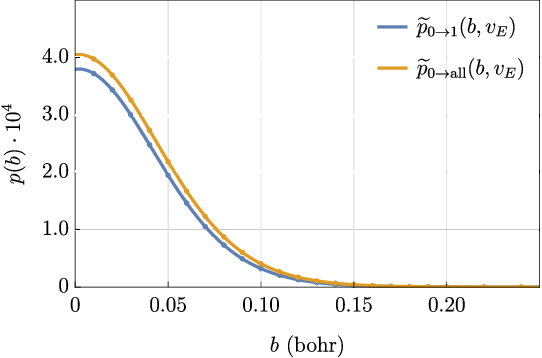} \hspace{1.0cm} & 
    \includegraphics[scale=0.86]{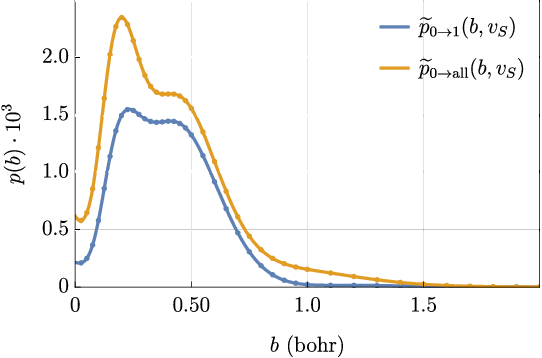} \\
    \end{tabular}
    \caption{Excitation probabilities $\widetilde{p}_{0\rightarrow 1}(b,v)$ and $\widetilde{p}_{0\rightarrow \mathrm{all}}(b,v)$ for the argon atom as a function of the impact parameter, $b$. The results for the gravitino velocities $v_E$ and $v_S$ are given in the left and right panel, respectively. All values are in atomic units. Notice a different scale on the vertical axis of both plots.}
    \label{fig:ar}
\end{figure*}

The fifth, and final, variable is the angle $\phi_v\in[0,2\pi]$ which rotates the path of the gravitino within a plane that passes through $\mathbf{s}_0$ and is perpendicular to the vector pointing from the origin to the point $\mathbf{s}_0$. One can show by direct calculation that the following vectors lie within this plane and are orthonormal:
\small
\begin{align*}
     &\mathbf{w_{1 \perp}} = 
     \frac{[1, \; 0, \; -x_0/z_0]}
     {\sqrt{1 + x_0^2/z_0^2}} \\
     &\mathbf{w_{2 \perp}} = 
   \frac{[  x_0y_0/z_0, \;
    -  x_0^2/z_0 - z_0, \;
    y_0 ]}{\sqrt{\frac{x_0^2}{z_0^2}(x_0^2 + y_0^2) + 2x_0^2 + y_0^2 + z_0^2}}
\end{align*}
\normalsize
While this choice of $\mathbf{w}_{1\perp}$ and $\mathbf{w}_{2\perp}$ is not unique, it is convenient in practice. Note that the definitions given above break down when $z_0=0$. However, in this special case one can set $\mathbf{w}_{1\perp} = [0, 0, 1]$ and $\mathbf{w}_{2\perp} = [- y_0, x_0, 0]$ without loss of generality. Finally, the last parameter specifying the path is
the angle $\phi_v$ is defined such that:
\begin{align}
    \hat{\mathbf{v}} = \frac{\mathbf{v}}{|\mathbf{v}|} = 
    \mathbf{w}_{1\perp} \cos\phi_v + 
    \mathbf{w}_{2\perp} \sin\phi_v.
\end{align}
The definition of the variables specifying the path is illustrated graphically in Fig.~\ref{fig:path}.

To ease the presentation of the results we define the excitation probability $\widetilde{p}_{0\rightarrow n}(b,v)$ averaged over the angles $\theta_s$, $\phi_s$, and $\phi_v$ according to the formula
\begin{align}
\label{p0nangle}
    \widetilde{p}_{0\rightarrow n}(b,v) = 
    \int_0^\pi d\theta_s\;\sin\theta_s \int_0^{2\pi} d\phi_s \int_0^{2\pi} d\phi_v\;
    p_{0\rightarrow n},
\end{align}
where the integrand is defined through Eq.~(\ref{p0n}). In the current implementation, this averaging is carried out numerically -- we use the Lebedev quadrature~\cite{lebedev76,lebedev77,lebedev99} with 302 points for integration over the solid angle $(\theta_s,\phi_s)$ and the trapezoidal rule with 30 points and equal grid spacing for integration over $\phi_v$.

In the last step, the excitation probabilities have to be averaged over the impact parameter, $b$. For simplicity, we assume that the scintillating medium in the inner detector consists solely of LAB with the bulk density $0.8650\,$g/cm$^3$ at $T=300\,$K~\cite{lombardi19}. The fluor and wavelength shifter molecules can be neglected in this analysis due to their low concentrations. Using the average molecular mass of LAB equal to $M_{\mathrm{LAB}}=241.52\,$g/mol, see Table~\ref{tab:lab-composition}, this leads to the density $\rho_{\mathrm{LAB}}=2.157\cdot10^{27}\,$m$^{-3}$ expressed as the number of molecules per unit volume. Similarly, we use $\rho_{\mathrm{Ar}}=2.136\cdot10^{28}\,$m$^{-3}$ for liquid argon. The distribution of molecules in the detector medium is assumed to be perfectly uniform. In the remainder of this section, we consider LAB molecules as an example, but analogous definitions hold also for liquid argon.

As the gravitino travels the distance $d$ through the inner detector, it encounters
\begin{align}
    dn_{\mathrm{LAB}^*} = 2\pi d\,\rho_{\mathrm{LAB}}\,b\,db,
\end{align}
LAB molecules with respect to which the gravitino path is characterized by the impact parameter in the interval $[b,b+db]$. Therefore, to find the total number of LAB molecules which get excited, $n_{\mathrm{LAB}^*}$, we have to evaluate the integral
\begin{align}
\begin{split}
    n_{\mathrm{LAB}^*} &= \int_0^\infty dn_{\mathrm{LAB}^*}\;\widetilde{p}_{0\rightarrow 1}(b,v) \\
    &=2\pi d \,\rho_{\mathrm{LAB}} \int_0^\infty db\;b\,\widetilde{p}_{0\rightarrow 1}(b,v),
\end{split}
\end{align}
where we restrict ourselves only to the first excited state of LAB. Further in the paper, we consider cross sections defined as
\begin{align}
\label{sigmab}
    \sigma_{\mathrm{LAB}^*}(v) = 2\pi \int_0^\infty db\;b\,\widetilde{p}_{0\rightarrow 1}(b,v),
\end{align}
such that $n_{\mathrm{LAB}^*} = d \,\rho_{\mathrm{LAB}}\,\sigma_{\mathrm{LAB}^*}(v)$. Note that the cross section $\sigma_{\mathrm{LAB}^*}(v)$ has units of m$^2$ and can be interpreted as the probability of exciting LAB to the first excited state per unit distance the gravitino travels and per single LAB molecule.

In the next two sections we report the calculated excitation probabilities and total cross sections for argon and model LAB molecule, while the corresponding results for water molecule are given in the Supplemental Material~\cite{supp}. On a technical note, the excitation energies and excited state wavefunctions required in $\widetilde{p}_{0\rightarrow 1}(b,v)$ were evaluated using the ADC(2) method~\cite{schirmer82,trofimov06,hattig02,dreuw15} using the aug-cc-pVXZ basis set family~\cite{dunning89,kendall92}. We used X=Q for argon atom while X=T was employed for the remaining systems. To reduce the cost of the calculations, density-fitting approximation of the two-electron integrals~\cite{whitten73,dunlap79,vahtras93,feyer93,rendell94,kendall97,weigend02} was invoked. The matching aug-cc-pVXZ-RIFIT basis sets optimized in Ref.~\cite{weigend98,weigend02b} were used for this purpose. The error resulting from the density-fitting approximation is negligible in the present context. 

\subsection{Excitation cross-sections: argon}
\label{subsec:cross-ar}

\begin{figure*}
    \centering
    \begin{tabular}{cc}
    \includegraphics[scale=0.85]{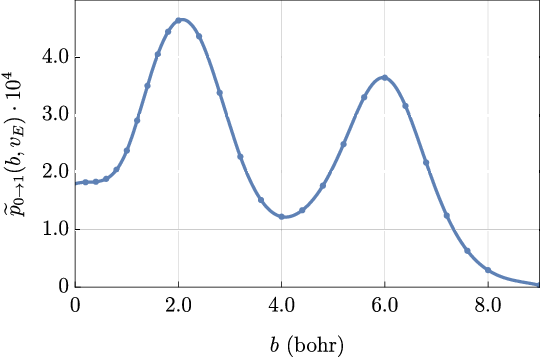} \hspace{1.0cm} & 
    \includegraphics[scale=0.85]{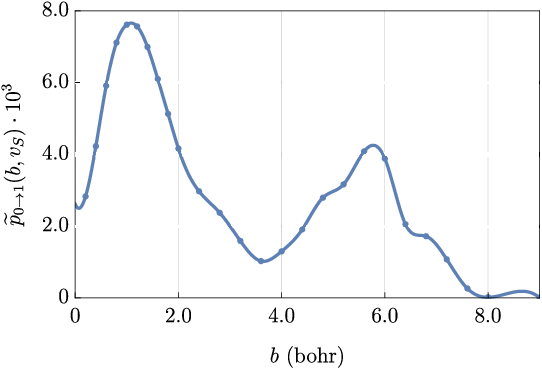} \\
    \end{tabular}
    \caption{Excitation probabilities $\widetilde{p}_{0\rightarrow 1}(b,v)$ for model LAB molecule as a function of the impact parameter, $b$. The results for the gravitino velocities $v_E$ and $v_S$ are given in the left and right panel, respectively. All values are in atomic units. Notice a different scale on the vertical axis of both plots.}
    \label{fig:lab}
\end{figure*}

We first consider liquid argon because of its possible relevance to the DUNE detector. The computations are considerably simpler than for the oil, as no elaborate path averaging procedure is required. For this case, excitation probabilities were evaluated for two gravitino velocities ($v_E$ and $v_S$). In the case of $v_E$, 32 values of $b$ were considered within the interval $b\in[0.0,0.50]$. For $v_S$ this grid is somewhat larger ($45$ points within interval $b\in[0.0,2.5]$) due to more rapid variation of the excitation probabilities as a function of $b$. As the path averaging is not necessary for argon, we calculated transition probabilities not only to the first excited state, i.e.~$\widetilde{p}_{0\rightarrow 1}(b,v)$, but also to all bound states supported by a given basis (with excitation energies lower than the ionization energy of the atom calculated by the same method). The total excitation probability is then obtained by summing the excitation probabilities $\widetilde{p}_{0\rightarrow n}(b,v)$ for all accessible $n$. We denote this quantity by the symbol $\widetilde{p}_{0\rightarrow \mathrm{all}}(b,v)$. The excitation probabilities $\widetilde{p}_{0\rightarrow 1}(b,v)$ and $\widetilde{p}_{0\rightarrow \mathrm{all}}(b,v)$ are given in Fig.~\ref{fig:ar} for the velocities $v_E$ and $v_S$, respectively. Clearly, the transition to the first excited state dominates, especially for the lower velocity of the gravitino. This suggests that the effects such as ionization are likely negligible.

In the calculation of the cross sections, we employ the total excitation probabilities $\widetilde{p}_{0\rightarrow \mathrm{all}}(b,v)$. In order to perform integration over the impact parameter $b$ according to Eq.~(\ref{sigmab}) we interpolate the results obtained on a grid using cubic $B$-spline method~\cite{numrep}. The integration in Eq.~(\ref{sigmab}) is then elementary and gives (in units of m$^2$)
\begin{align}\label{tauArgon}
\begin{split}
    &\sigma_{\mathrm{Ar}^*}(v_E) = 6.84\cdot10^{-27}, \\
    &\sigma_{\mathrm{Ar}^*}(v_S) = 1.73\cdot10^{-24},
\end{split}
\end{align}
for the velocities $v_E$ and $v_S$, respectively. From the point of view of simulations described in the next section, it is useful to define a related quantity, namely the mean time between the excitation events as the gravitino travels through the scintillating medium. This characteristic time can be calculated from the formula
\begin{align}
    \tau_{\mathrm{Ar}^*}(v) = 
    \frac{1}{v\,\sigma_{\mathrm{Ar}^*}(v)\rho_{\mathrm{Ar}}},
\end{align}
and in the present case it equals 
\begin{align}
\begin{split}
    &\tau_{\mathrm{Ar}^*}(v_E) = 228\,{\rm ns} \\
    &\tau_{\mathrm{Ar}^*}(v_S) = 0.12\,{\rm ns}.
\end{split}
\label{Arcs}
\end{align}
The argon-based experiments performed up to now had a relatively small size.  If the particles in question
actually exist, the absence of any signal so far in these 
experiments is therefore most likely due to the much smaller surface area of existing DM detectors
and possibly to the fact that these experiments did not expressly 
look for effects with a longer duration induced by atomic excitations, 
rather focusing on nuclear-recoil-induced events may have prevented 
a discovery even if a supermassive gravitino had actually passed through the detector.
However, as we can see from (\ref{Arcs}) the number of photons produced by excitations in liquid argon can be significant and there is in principle a potential for discovery in the bigger detectors, such as the one with 20 t of liquid argon at Gran Sasso \cite{Ar20t} and definitely in the extremely large DUNE \cite{DUNE}.

\subsection{Excitation cross-sections: LAB scintillator}
\label{subsec:cross-lab}

Analogous calculations were performed for the LAB model molecule for velocities $v_E$ and $v_S$, and this is the main result we want to report in this work. We used the a grid of $b$ with spacing $0.2$ within the interval $(0.0,2.0)$ (where the excitation probabilities are the largest), and with spacing $0.4$ within the interval $(2.0,9.0)$. Beyond the point $b=9.0$, the excitation probabilities decay rapidly and are negligible for both velocities. The results for the model LAB molecule are illustrated in Fig.~\ref{fig:lab}.

For integration over $b$, we used the same approach as in the case of argon atom. This leads to the averaged cross sections (in m$^2$)
\begin{align}
\begin{split}
    &\sigma_{\mathrm{LAB}^*}(v_E) = 5.62\cdot10^{-23}, \\
    &\sigma_{\mathrm{LAB}^*}(v_S) = 5.74\cdot10^{-22},
\end{split}
\end{align}
and characteristic times
\begin{align}  
\begin{split}
    &\tau_{\mathrm{LAB}^*}(v_E) = 0.27\,{\rm ns} \\
    &\tau_{\mathrm{LAB}^*}(v_S) = 0.0035\,{\rm ns}.
\end{split}
\label{labcs}
\end{align}
These results are employed in the simulations described in the next section. Note that the excitation probabilities reported in this work never exceed a few parts per thousand, even at the peak $b$ and in the case of the larger velocity $v_S$. This suggests that the excitation process is indeed perturbative in nature and the adopted theoretical framework is justified. 
Note that by comparing the results for argon and LAB, i.e. Eqs.~(\ref{Arcs})~and~(\ref{labcs}), it is immediately evident that the photon production rates for LAB are (much) higher than the ones for argon, which is another significant advantage of JUNO in comparison with previous experiments.

\section{Simulations}

In this section we provide some examples of simulations that show how  the photons produced by a gravitino traversing the detector would be registered by the JUNO PMTs. As was shown in the previous section, the excitations of the scintillator oil with its admixtures of PPO and Bis-MSB on average produce a large number of photons per nanosecond along the track of the gravitino which strongly depends on the velocity of the gravitino. Specifically, 
we simulate the events for two velocities; galactic one assumed to be 230 km/s ($\beta_S\approx 7.7\cdot 10^{-4}$) 
and 29.8 km/s ($\beta_E\approx 10^{-4}$). For these exemplary values  we have produced simulations 
with the following assumptions, based on the information provided in Ref.~\cite{JUNO}:
\begin{itemize}
\item
The photons produced by a passing gravitino along the track
are emitted in random directions every 0.0035 ns (for $\beta_S$) or 0.27 ns 
(for $\beta_E$) as shown in Eq.~(\ref{labcs});
\item
the overall efficiency of the photomultipliers is 30.1\%;
\item 
the area coverage by large (20'') PMTs is 75\%;
\item
the attenuation coefficient of photons in the oil is given by $\exp(-l/l_a)$ with  $l_a$=27 m;
\item
the radioactive $^{14}$C has a concentration $^{14}$C/$^{12}$C of at most $2\cdot 10^{-17}$ \cite{Oberauer};
\item 
dark count rates are counted separately for two types of 20'' PMTs used in the central detector, namely 
Hamamatsu (5000 PMTs) and NNTV (12612 PMTs) \cite{JUNOtesting}.
\end{itemize} 
Other sources of photons, like $^{238}$U chain, $^{232}$Th chain,  $^{210}$Pb chain, $^{85}$Kr contaminations as well as external sources, like radiation from surrounding rocks, are not included in the simulations. The flashes from antineutrinos and muons are very intense but very short-lived, and should be subtracted if they occur during the passage of a gravitino.

Based on these assumptions we get the average number of photons coming from the passage of gravitino and detected by large PMTs:
\bea
\label{photonrate}
0.301\cdot 0.75 \cdot 0.9 /(0.0035\ {\rm ns})&\sim& 58 /{\rm ns}
\;\;\;\;\,\mbox{for}\;\;\beta_S,\;\;\;\;\;\; \\
0.301\cdot 0.75 \cdot 0.9/(0.27\ {\rm ns})&\sim& 0.75 /{\rm ns}
\;\;\mbox{for}\;\;\beta_E,\;\;\;\;\;\;
\eea
where the first factor in  the numerator corresponds to the PMT efficiencies, the second to the area coverage and the third to the approximate contribution from the attenuation in the oil. The time of passage for
a gravitino crossing the detector along the diagonal would be 160 $\mu$s for $\beta=7.7\cdot 10^{-4}$ and 1260 $\mu$s for $\beta=10^{-4}$ so the number of photons detected should be $\sim$ 3.5 million in the first case and almost a million in the second. For tracks close to the edge, like the one we have chosen for simulations as described below, these numbers will be proportionally smaller but the relative importance of both $^{14}$C decays and the dark count rates is the same in all cases.

\begin{figure*}[t]
    \centering
    \includegraphics[scale=1.00]{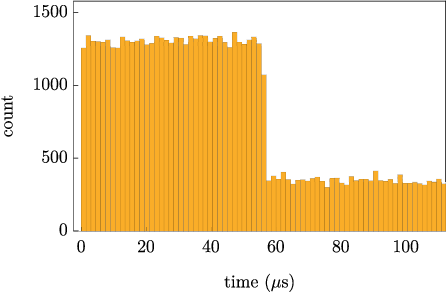}\hspace{1.0cm}
    \includegraphics[scale=1.02]{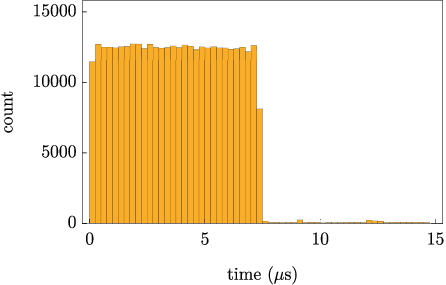}
    \caption{
    Number of detected photons as a function of time for velocities $v_E$ (left panel) and $v_S$ (right panel).}
    \label{fig:histogram}
\end{figure*}

\subsection{Effects of $^{14}$C contamination}

We have taken into account the decays $^{14}\mathrm{C} \rightarrow \,^{14}\mathrm{N} + e^- + \bar\nu$, assuming a relative $^{14}$C/$^{12}$C concentration of $2\cdot~10^{-17}$ (representing an {\em upper} bound \cite{Oberauer}), and $^{14}$C half-life of $1.2\cdot 10^{11}$~s. This gives, on average, 0.143 decays per microsecond inside the detector, assuming the mass of oil equal to $2\cdot 10^7$~kg.  The actual number of photons depends on 
the fraction $x$ of the energy carried away by the electron (with a maximal value of 156 keV), for which we use the beta distribution
\beq
P(x;p,q) \,=\, \frac{x^{p-1}(1-x)^{q-1}}{{\rm B}(p,q)} \qquad (0\leq x \leq 1),
\eeq
where the energy of the electron is $x\cdot$156 keV. Inspection of the distribution curve given in \cite{C14spectrum} shows that the best fit for the $^{14}$C decays is achieved with the parameter values $p = 1.2$ and $q =2.7$  and the number of photons emitted is assumed to be ten photons per every keV of electron's energy.  Assuming that the $^{14}$C decay gives on average 49 keV (i.e. 490 photons) we have the following approximate number of photons per microsecond detected by large PMTs 
\beq
0.143\cdot 490\cdot 0.301\cdot 0.75 /\mu {\rm s}\sim 0.016 /{\rm ns}.
\eeq
As we see in Eq.~(\ref{photonrate}) this value is roughly 3000 times smaller than the photon detection rate from $v_S$ gravitinos and 40 times smaller than the photon detection rate for $v_E$ gravitinos, so even in the 
second case they do not influence the simulations in any significant way (but are included in the 
simulations nevertheless).

\subsection{Dark count rate}

There are two different 20'' PMTs used in the JUNO Central Detector (CD) which cover about 
75\% of the total area: HPK (Hamamatsu, 5000 PMTs) and  NNVT (12612 PMTs)  (all the data in 
this section are taken from \cite{JUNOtesting}). There are also 25600 small 3'' PMTs, whose combined area is only a few percent of the total area, so they are not included in the simulations. The photon detection efficiency (PDE) of both types is similar and the average PDE for the PMTs in CD is 30.1\%. The dark count rates (DCRs) are however significantly different -- HPK PMTs have DCRs narrowly centered about 15 kHz, while NNVT PMTs have a broad distribution with an average 50 kHz (potted 31 kHz). These numbers were measured on the surface and normalized to 12 hours in darkness. Therefore one can expect the actual DCRs to be smaller after these PMTs are mounted in the CD, both because of a much longer period of darkness and because of a much smaller muon flux at the experiment site. We assume that these numbers are 8 kHz (HPK) and 15 kHz (NNVT), respectively. It could be a priori possible to count only photons registered by the PMTs with lower DCR (HPK) that cover approximately one quarter of the full solid angle and compare it with the full count (HPK+NNVT) to determine the influence of DCR 
on the reconstruction of the event. 

Taking this into account we get the dark count every 4.4 ns, {\em i.e.} the rate of dark counts is  0.23/ns
if we take all (large) PMTs (and every 25 ns if we take only HPK PMTs but then we have only 40\% of the previous area). This does not present any problem for $\beta=7.7\cdot 10^{-4}$. However, for $\beta=10^{-4}$ it is a significant fraction of the gravitino induced photons. However, as will be shown below, even in this case it is unambiguous to detect the passage of the gravitino and distinguish it from the statistical fluctuations, given the very large number of photons and using the statistical $F$ distribution, we have extremely high level of significance. A passage of a gravitino can thus be proven practically with certainty.

\subsection{Results of simulation}

We have found that the signature for 'fast' gravitinos ($\beta_S$) is absolutely unambiguous, due to the very large number of photons emitted in comparison with any background.  We display here the simulation results for the cases of both slow gravitinos ($\beta_E$) and fast gravitinos ($\beta_S$) with a relatively short track close to the edge of the ball. In all other cases the visibility of the track is much more pronounced.

The track we have chosen to visualize is  close to the edge of the ball, directed vertically upwards and has the following entry and exit points
\bea
(x_i,y_i,z_i)&=&(16.47,1.65,-3.86)\ {\rm m}\nn\\
(x_f,y_f,z_f)&=&(16.77,1.68,-2.19)\ {\rm m}
\label{infpositions}
\eea
(the radius is assumed to be equal to 17 m). From our point of view this represents
a kind of `worst case',  in the sense that with other parameters the track is in general more visible. Indeed, we
have simulated other tracks, closer to the center of the ball and at different angles and the results were very similar when the lengths of the tracks (and respective times of passages) were appropriately rescaled with respect to the one chosen here.

We discuss first the slow gravitino ($\beta =10^{-4}$).
We start to count the time when the gravitino enters the ball, then, after it exits the ball (in 56 $\mu$s) we continue the simulation up to 112 $\mu$s to see the difference between signal plus background in the first 56 $\mu$s and only background in the next 56 $\mu$s.

\begin{figure}[t]
    \centering
    \hspace{-0.5cm}\includegraphics[scale=0.75]{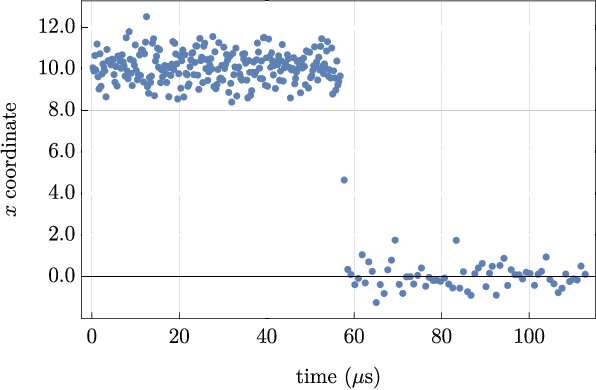} \\\vspace{0.5cm}
    \hspace{-0.5cm}\includegraphics[scale=0.75]{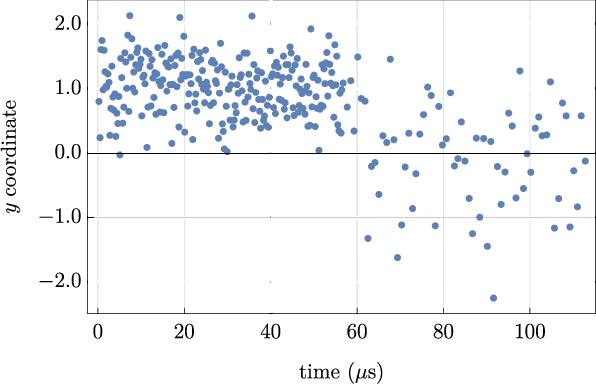} \\\vspace{0.5cm}
    \hspace{-0.5cm}\includegraphics[scale=0.75]{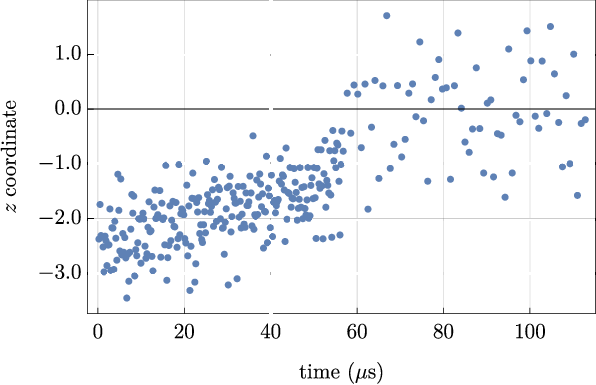}
    \caption{Averaged (over 200 subsequent firings) $x$, $y$, $z$ positions of the firing PMTs as a function of time (gravitino velocity $v_E$).}
    \label{fig:coordinates}
\end{figure}

In Fig.~\ref{fig:histogram} on the left we display the simulated number of detected photons as a function of time, taking into account all the assumptions listed above. As we can see the passage of gravitino can be clearly and unambiguously seen as being very different for the first 56 $\mu$s and the next 56 $\mu$s. No background or other particle can mimic such a signal.

\begin{figure}[t]
    \centering
    \hspace{-0.5cm}\includegraphics[scale=0.75]{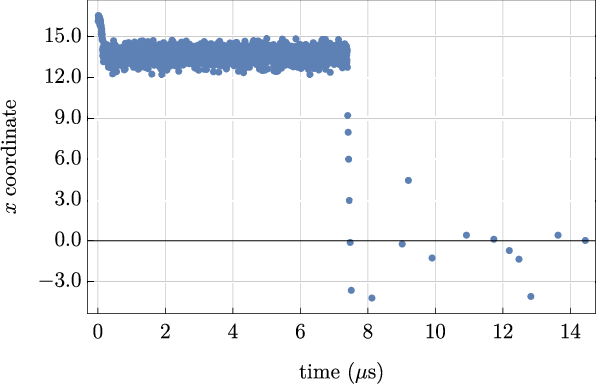} \\\vspace{0.5cm}
    \hspace{-0.5cm}\includegraphics[scale=0.75]{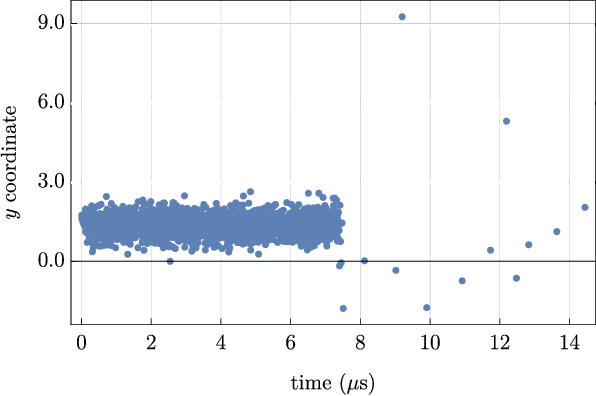} \\\vspace{0.5cm}
    \hspace{-0.5cm}\includegraphics[scale=0.75]{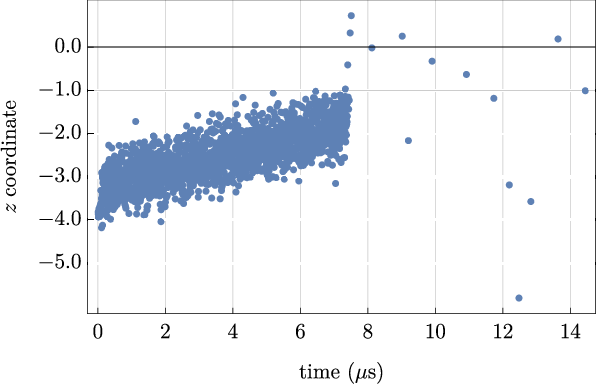}
    \caption{Averaged (over 200 subsequent firings) $x$, $y$, $z$ positions of the firing PMTs as a function of time (gravitino velocity $v_S$).}
    \label{fig:coordinates2}
\end{figure}

The track can then be reconstructed as follows. Since photons are emitted in all directions, to approximately reproduce the track it is reasonable to average over the positions of the firing photomultipliers -- we have chosen to average over 200 consecutive firings. Fig.~\ref{fig:coordinates} presents the results of the simulation in an attempt to reconstruct the track. The horizontal axis is in microseconds and the vertical axis in meters showing on three figures average positions $x$, $y$ and $z$ of firing photomultipliers as a function of time of firing (they are the times of arrival, not the times of emission). As we can see comparing with (\ref{infpositions}), even in this unfavorable case (slow gravitino with smaller ratio signal/background than for the fast ones) we can approximately reproduce the track.

Next we discuss the case of fast gravitinos ($\beta =7.7\cdot 10^{-4}$). We start to count the time when the gravitino enters the ball, then, after it exits the ball (in 7 $\mu$s) we continue the simulation up to 14 $\mu$s to see the difference between signal plus background in the first 7 $\mu$s and only background (almost invisible in comparison) in the next 7 $\mu$s. In Fig.~\ref{fig:histogram} on the right we display the simulated number of detected photons as a function of time. As we can see the passage of a gravitino can be even more clearly and unambiguously detected than before as being very different for the first 7 $\mu$s and the next 7 $\mu$s. We approximately reproduce the track after averaging, as before, over 200 consecutive firings and Fig.~\ref{fig:coordinates2} presents the results of the simulation. When compared with (\ref{infpositions}) we see that the track can be well reproduced as a function of time with several meters of precision.

\section{Final comments}
\label{sec:discussion}

Our calculation vividly illustrates why and how JUNO sits in a `sweet spot' 
predestined to search for supermassive charged gravitinos. This is true not only for 
the most favorable case of galactic speeds $v\sim v_S$, for which the number of photons
produced along the track is very large, but also for slow, `sun-bound'  gravitinos with $v\sim v_E$. 
As the simulations show, any passage of a gravitino, either slow or fast, close to the center or to the edge, should produce a clear and unmistakable signal if the electronics, triggering system and the software are set to
take into account the possibility of long and continuous detection of photons over many microseconds. Similar remarks can be made what concerns a possible detection of charged gravitinos in DUNE.
Finally, we would like to reiterate that the excitation cross sections for LAB reported in this work are conservative lower bounds, taking into account only the $S_0\rightarrow S_1$ excitation process. In subsequent paper, we shall consider other competing processes initiated by the passing gravitino, such as ionization or capture of an electron. The impact of these processes on the strength of the signal will be studied, improving the current lower bound.

\vspace{1cm}

\begin{acknowledgments}
We would like to thank Lothar Oberauer, S{\l}awomir  Wronka and Agnieszka Zalewska for helpful discussions. K.A.~Meissner thanks Albert Einstein Institute for hospitality and support. K.A.~Meissner was partially supported by the Polish National Science Center grant UMO-2020/39/B/ST2/01279. The work of  H.~Nicolai has received funding from the European Research Council (ERC) under the  European Union's Horizon 2020 research and innovation programme (grant agreement No 740209). We gratefully acknowledge Poland's high-performance Infrastructure PLGrid (HPC Centers: ACK Cyfronet AGH, PCSS, CI TASK, WCSS) for providing computer facilities and support within computational grant PLG/2023/016599.
\end{acknowledgments}

\bibliography{gravitino}

\end{document}